\documentclass[aps,prd,twocolumns,eqsecnum,preprintnumbers,showpacs,amsmath,amssymb]
{revtex4}

\usepackage{graphicx}
\usepackage{dcolumn}
\usepackage{bm}

\begin{document}

\title{RENORMALIZATION OF THE MASS GAP }

\author{V. Gogokhia}
\email[]{gogohia@rmki.kfki.hu}

\affiliation{HAS, CRIP, RMKI, Depart. Theor. Phys., Budapest 114,
P.O.B. 49, H-1525, Hungary}

\date{\today}
\begin{abstract}
The full gluon propagator relevant for the description of the
truly non-perturbative QCD dynamics, the so-called intrinsically
non-perturbative gluon propagator has been derived in our previous
work. It explicitly depends on the regularized mass gap, which
dominates its structure at small gluon momentum. It is
automatically transversal in a gauge invariant way. It is
characterized by the presence of severe infrared singularities at
small gluon momentum, so the gluons remain massless, and this does no
depend on the gauge choice. In this paper we have shown how
precisely the renormalization program for the regularized mass gap
should be performed. We have also shown how precisely severe
infrared singularities should be correctly treated. This allowed
to analytically formulate the exact and gauge-invariant criteria
of gluon and quark confinement. After the renormalization program
is completed, one can derive the gluon propagator applicable for
the calculation of physical observables, processes, etc., in
low-energy QCD from first principles.

\end{abstract}



\maketitle

\section{Introduction}

In our previous work \cite{1} the mass gap responsible for the
non-perturbative dynamics in QCD has been introduced through the
definition of the subtracted full gluon self-energy. So it is
defined as the value of the regularized full gluon self-energy at
some finite point. The mass gap is mainly generated by the
nonlinear interaction of massless gluon modes. We have explicitly
shown that QCD is the color gauge invariant theory at non-zero
mass gap as well. All this allows one to establish the structure
of the full gluon propagator in the explicit presence of the mass
gap \cite{1}. In this case, the two independent general types of
the formal solutions for the regularized full gluon propagator
have been found \cite{2}. No
truncations/approximations/assumptions, as well as no special
gauge choice are made for the regularized skeleton loop integrals,
contributing to the full gluon self-energy. The so-called massive
solution, which leads to an effective gluon mass has been
established. The general nonlinear iteration solution for the full
gluon propagator depends explicitly on the mass gap. It is always
severely singular in the $q^2 \rightarrow 0$ limit, so the gluons
remain massless, and this does not depend on the gauge choice.
However, we have argued that only its intrinsically
non-perturbative (INP) part, defined by the subtraction of all the
types of the perturbative theory (PT) contributions
("contaminations") from the general nonlinear iteration solution,
is interesting for confinement. In this way it becomes
automatically transversal in a gauge invariant way and excludes
the free gluons from the theory. This is important for correct
understanding of the color confinement mechanism, as underlined in
our previous works \cite{1,2}.

The INP gluon propagator is to be used for the numerical
calculations of physical observables, processes, etc. in
low-energy QCD from first principles. But before the two important
problems remain to solve. The first problem is how to perform the
renormalization program for the regularized mass gap $\Delta^2
\equiv \Delta^2(\lambda, \alpha, \xi, g^2)$, and to see whether
the mass gap survives it or not. The second problem is how to
treat correctly severe infrared (IR) singularities $(q^2)^{-2-k},
\ k=0,1,2,3,...$ inevitably present in this solution (let us
remind that the signature is Euclidean). Just these problems are
to be addressed and solved in this paper.

\section{The INP gluon propagator}

The full gluon propagator, which is relevant for the description
of the truly NP QCD dynamics, the so-called INP gluon propagator
derived in Ref. \cite{2} is as follows:

\begin{equation}
D^{INP}_{\mu\nu}(q) = i T_{\mu\nu}(q) d^{INP}(q^2) {1 \over q^2} =
i T_{\mu\nu}(q) {\Delta^2 \over (q^2)^2} L(q^2),
\end{equation}
and

\begin{equation}
L(q^2) \equiv L(q^2; \Delta^2) = \sum_{k=0}^{\infty} \Bigl(
{\Delta^2 \over q^2} \Bigr)^k \Phi_k =\sum_{k=0}^{\infty} \Bigl(
{\Delta^2 \over q^2} \Bigr)^k \sum_{m=0}^{\infty} \Phi_{km},
\end{equation}
where $\Delta^2 \equiv \Delta^2(\lambda, \alpha, \xi, g^2)$ and
the residues $\Phi_k \equiv \Phi_k(\lambda, \alpha, \xi, g^2)$ and
thus $\Phi_{km} \equiv \Phi_{km}(\lambda, \alpha, \xi, g^2)$ as
well.

Let us recall some interesting features of the INP gluon propagator
(2.1) \cite{2}. First of all, it depends only on the transversal
degrees of freedom of gauge bosons. Also, its functional
dependence is uniquely fixed up to the expressions for the
residues $\Phi_k$, and it is valid in the whole energy/momentum
range. It explicitly depends on the mass gap, so that when it
formally goes to zero (the formal PT $\Delta^2=0$ limit) this
solution vanishes. Also, it is characterized by the explicit
presence of severe (i.e., NP) IR singularities $(q^2)^{-2-k}, \
k=0,1,2,3,...$ only. The sum over $m$ indicates that an infinite
number of iterations (all iterations) invokes each severe IR
singularity labeled by $k$. Apart from the structure $( \Delta^2
/ (q^2)^2$ it is nothing but the corresponding Laurent expansion.
This solution is completely free of all the types of the PT
contributions ("contaminations"), and thus is exactly and uniquely
separated from the PT gluon propagator, indeed. Due to the regular
dependence on the mass gap, it dominates over the free gluon
propagator in the deep IR ($q^2 \rightarrow 0$) limit and it is
suppressed in the deep ultraviolet (UV) ($q^2 \rightarrow \infty$)
limit (let us remind that the free gluon propagator behaves like
$1 / q^2$ in the whole momentum range). In the rest of this paper
we will omit the superscript "INP", for simplicity. We will
restore it if it will be necessary for clarity of our discussion.

\section{The general multiplicative renormalization program}

One of the remarkable features of the solution (2.1)-(2.2) is that
its asymptotic at infinity ($q^2 \rightarrow \infty$) is to be
determined by its $ \Delta^2 /(q^2)^2$ structure only, since all
other terms in its expansion are suppressed in this limit. It is
well known that such a behavior at infinity is not dangerous for
the renormalizability of QCD \cite{3,4,5,6}. However, the
regularized mass gap itself is of the NP origin. So its
renormalization in order to get finite result in the $\lambda
\rightarrow \infty$ limit cannot be done in standard way. Another
main problem, closely connected to the mass gap, is its violent
structure in the deep IR region ($q^2 \rightarrow 0$).
Fortunately, there exist two mathematical theories which are of a
great help in this case. The distribution theory (DT) \cite{7},
which should be complemented by the dimensional regularization
method (DRM) \cite{8} (some important issues of these theories are
present and discussed in appendix A in more detail).

For further purpose, let us present our solution (2.1) as follows:

\begin{equation}
D_{\mu\nu}(q) = i T_{\mu\nu}(q) \sum_{k=0}^{\infty}
(\Delta^2)^{k+1} (q^2)^{-2-k} \Phi_k.
\end{equation}
As it has been emphasized in appendix A, a special regularization
expansion is to be used in order to deal with the severe IR
singularities $(q^2)^{-2-k}$, since standard methods fail to
control them. If $q^2$ is an independent loop variable, then the
corresponding dimensionally regularized expansion is given in
Eq.~(A10), namely

\begin{equation}
(q^2)^{-2-k}  = {1 \over \epsilon} \Bigr[ a(k) [\delta^4(q)]^{(k)}
+ O_k(\epsilon) \Bigl], \quad \epsilon \rightarrow 0^+.
\end{equation}
Substituting it into the previous expansion, one obtains

\begin{equation}
D_{\mu\nu}(q) = i T_{\mu\nu}(q){ 1 \over \epsilon}
\sum_{k=0}^{\infty} (\Delta^2)^{k+1} \Phi_k \Bigr[
a(k)[\delta^4(q)]^{(k)} + O_k(\epsilon) \Bigl], \quad \epsilon
\rightarrow 0^+,
\end{equation}
so instead of the Laurent expansion (2.1) we obtain dimensionally
regularized (in powers of $\epsilon$) expansion for the INP gluon
propagator. In the presence of the IR regularization parameter
$\epsilon$ all the quantities and parameters depend, in general,
on it. For further purpose it is convenient to introduce the
following short-hand notations, namely $\Delta^2 \equiv
\Delta^2(\lambda, \epsilon)$ and $\Phi_k \equiv \Phi_k (\lambda,
\epsilon)$ and omit the dependence on the other parameters, for
simplicity. Evidently, the renormalization of the mass gap only is
what that matters, since $\Phi_k (\lambda, \epsilon)$ is the
dimensionless quantity (see below).

So let us define the UV and IR renormalized mass gap as follows:

\begin{equation}
\Delta^2(\lambda; \epsilon) = Z(\lambda; \epsilon)
\bar{\Delta}^2_R, \quad \lambda \rightarrow \infty, \quad \epsilon
\rightarrow 0^+,
\end{equation}
where $Z \equiv Z(\lambda; \epsilon)$ is the multiplicative
renormalization (MR) constant of the mass gap. Here and everywhere
below all the quantities with sub- or superscript "R" denote the
quantities which exist in the $\lambda \rightarrow \infty$ limit,
i.e., by definition, they are UV renormalized (and their
dependence on other parameters is to be neglected, apart from the
coupling constant, which can be "running" itself, see section V
below). At the same time, all the quantities with bar are IR
renormalized, i.e., by definition, they exist as $\epsilon
\rightarrow 0^+$. That is why here and everywhere below in all the
renormalized quantities the dependence on $\lambda$ and $\epsilon$
is not shown explicitly, as well as on other possible parameters.
The factorization of the corresponding MR constants in order to
get the finite results separately in the $\lambda \rightarrow
\infty$ and $\epsilon \rightarrow 0^+$ limits is a particular case
of the general MR program in order to remove the dependence on
$\lambda$ and $\epsilon$ in their corresponding limits. The NP
(quadratic) UV divergences present in the regularized mass gap are
in close intrinsic link with the NP IR singularities, which both
are present in the INP QCD gluon propagator (2.1).

Substituting the relation (3.4) into the previous expansion (3.3),
one arrives at

\begin{equation}
D_{\mu\nu}(q) = i T_{\mu\nu}(q){ 1 \over \epsilon}
\sum_{k=0}^{\infty} (\bar{\Delta}^2_R)^{k+1} Z^{k+1}(\lambda;
\epsilon) \Phi_k(\lambda; \epsilon) \Bigr[ a(k)
[\delta^4(q)]^{(k)} + O_k(\epsilon) \Bigl], \quad \epsilon
\rightarrow 0^+.
\end{equation}
The only way to remove the pole $ 1 / \epsilon$ from the previous
expansion and get the final results after the removal the
dependence on $\lambda$ and $\epsilon$ is to put

\begin{equation}
Z^{k+1}(\lambda; \epsilon) \Phi_k(\lambda; \epsilon) = \epsilon
\bar{\Omega}_k^R, \quad \epsilon \rightarrow 0^+, \quad \lambda
\rightarrow \infty, \quad k=0,1,2,3,....,
\end{equation}
since all the NP IR singularities labeled by $k$ are independent
from each other. This is the general convergence condition which
makes the gluon propagator free of $\lambda$ and $\epsilon$ in
their corresponding limits.

So the previous expansion becomes

\begin{equation}
\bar{D}^R_{\mu\nu}(q) = i T_{\mu\nu}(q) \sum_{k=0}^{\infty}
(\bar{\Delta}^2_R)^{k+1} \bar{\Omega}_k^R \Bigr[ a(k)
[\delta^4(q)]^{(k)} + O_k(\epsilon) \Bigl], \quad \epsilon
\rightarrow 0^+.
\end{equation}
Through the chain of the relations $ \bar{\Omega}_k^R =
[\bar{\Psi}^R_k]^{k+1}$, $\bar{\Delta}^2_R \bar{\Psi}^R_k =
(\bar{\Delta}^2_R)_k$, one obtains

\begin{equation}
\bar{D}^R_{\mu\nu}(q) = i T_{\mu\nu}(q) \sum_{k=0}^{\infty}
[(\bar{\Delta}^2_R)_k]^{k+1} \Bigr[ a(k) [\delta^4(q)]^{(k)} +
O_k(\epsilon) \Bigl], \quad \epsilon \rightarrow 0^+,
\end{equation}
which shows that one can start from the $k$-dependent regularized
mass gap in Eq.~(3.3), but, nevertheless, coming to the same final
Eq.~(3.8), containing the $k$-dependent renormalized mass gap.

The renormalization of the mass gap itself has been defined by the
relation (3.4). Its MR constant $Z$ remains undetermined, but this
is not already the problem, since the renormalized gluon
propagator (3.7) depends on the renormalized mass gap
$\bar{\Delta}^2_R$. So we consider it as the physical mass gap
within our approach. Precisely this quantity should be positive,
finite, gauge- and $\alpha$-independent, etc., it should exist
when $\lambda \rightarrow \infty$ and $\epsilon \rightarrow 0^+$.
The renormalization of the mass gap is an example of the NP MR
program. We were able to accumulate all the quadratic divergences
into the renormalization of the mass gap. Due to the
renormalization the quadratic divergences parameterized as the
regularized mass gap $\Delta^2$ may be absorbed in its
re-definition leading thus finally to the physical mass gap.

Concluding, since all the parameters in the expansion (3.7) are
expressed in the IR renormalized quantities, we can go to the
$\epsilon \rightarrow 0^+$ limit, without encountering any
problems in this limit now. Also, we have found of no practical
use to introduce the MR constant $Z_3$ for the gluon propagator
itself separately from that of the mass gap (3.4).

\subsection{The general criterion of gluon confinement}

We are now in the position to analytically formulate the general
criterion of gluon confinement. It will be instructive to
analytically formulate the general criterion of quark confinement
as well. Let us begin with the former one.

The dimensionally renormalized (within the DT complemented by the
DRM) expression for the relevant gluon propagator in INP QCD is
Eq.~(3.7). Substituting back into this equation the expansion
(3.2), one obtains

\begin{equation}
D^R_{\mu\nu}(q) = \epsilon \times i T_{\mu\nu}(q)
\sum_{k=0}^{\infty} (\bar{\Delta}^2_R)^{k+1} \bar{\Omega}_k^R
(q^2)^{-2-k}, \quad \epsilon \rightarrow 0^+.
\end{equation}
Let us underline that this is the general expression for the
renormalized gluon propagator, since it does not depend on whether
the gluon momentum is independent skeleton loop variable or not.
Due to the distribution nature of the NP IR singularities
$(q^2)^{-2-k}, \ k=0,1,2,3....$ (see appendix A as well), the two
principally different cases should be separately considered.

{\bf 1.} If the gluon momentum $q$ is an independent skeleton loop
variable, then, as emphasized repeatedly above, the initial
$(q^2)^{-2-k}$ NP IR singularities should be regularized with the
help of the expansion (3.2). Finally one arrives at Eq.~(3.7) in
the $\epsilon \rightarrow 0^+$ limit, as it should be.

Let us note in advance that beyond the one-loop skeleton integrals
the analysis should be done in a more sophisticated way, otherwise
the appearance of the product of at least two $\delta$ functions
at the same point is possible. However this product is not defined
in the DT \cite{7}. So in the multi-loop skeleton diagrams instead
of the $\delta$ functions in the residues their derivatives may
appear (see Ref. \cite{7} and appendix A in this paper).  They
should be treated in the sense of the DT. Fortunately, as
mentioned in appendix A, the IR renormalization of the theory is
not undermined, since a pole in $\epsilon$ is always a simple pole
$1 / \epsilon$ for each independent skeleton loop variable (see
the dimensionally regularized expansion (A10)). That is why the
starting expression for  the relevant gluon propagator in INP QCD
is always the general expression (3.9). It makes it possible to
perform formal algebraic operations (first of all multiplication)
on the corresponding Laurent expansions. The final product is
again the Laurent expansion. This allows one to use the
dimensionally regularized expansion (A10), and thus to avoid the
multiplication of the $\delta$-functions at the same point, if the
starting expansion will be expansion (3.7) for each gluon
propagator.

{\bf 2.} The necessary and sufficient conditions for gluon
confinement are:

{\bf (i).} If, however, the gluon momentum $q$ is not a loop
variable (i.e., it is external momentum), then the initial
$(q^2)^{-2-k}$ NP IR singularities cannot be treated as
distributions, i.e., the regularization expansion (3.2) is not the
case to be used. The functions $(q^2)^{-2-k}$ are the standard
ones, and the relevant gluon propagator (3.9) vanishes as
$\epsilon$ goes to zero, i.e.,

\begin{equation}
D^R_{\mu\nu}(q) = \epsilon \times i T_{\mu\nu}(q)
\sum_{k=0}^{\infty} [(\bar{\Delta}^2_R)_k]^{k+1} (q^2)^{-2-k} \sim
\epsilon, \quad \epsilon \rightarrow 0^+.
\end{equation}
It is worth emphasizing that the final $\epsilon \rightarrow 0^+$
limit is permitted to take only after expressing all the Green's
functions, parameters and mass gap in terms of their IR
renormalized counterparts because they, by definition, exist in
this limit. This behavior is gauge-invariant, does not depend on
any truncations/approximations, etc., and thus it is a general
one. It prevents the transversal dressed gluons to appear in
asymptotic states, so color dressed gluons can never be isolated.
This is the {\bf first necessary} condition for gluon confinement.

{\bf (ii).} The {\bf second sufficient} condition of gluon
confinement is the absence of the free gluons in the corresponding
theory. Just such a theory has been formulated in our previous
works \cite{1,2}, namely INP QCD. Let us remind that its full gluon propagator (2.1)
has no free gluon propagator limit when the interaction is to be switched off ($\Delta^2=0$).
We argued that it should be used for the calculations of physical observables, processes in
low-energy QCD from first principles. We consider both the
suppression of the colored dressed gluons at large distances and the absence of the
free colored gluons in the theory as the exact criterion of
gluon confinement (for its initial formulation see Ref.
\cite{9}).

\subsection{The general criterion of quark confinement}

It is instructive to formulate here the quark confinement
criterion in advance as well. It consists of the two independent
conditions.

{\bf 1. The first necessary condition}, formulated at the
fundamental (microscopic) quark-gluon level, is the absence of the
pole-type singularities in the quark Green's function at any gauge
on the real axe at some finite point in the complex momentum
plane, i.e.,

\begin{equation}
S(p) \neq { Z_2 \over \hat p - m_{ph}},
\end{equation}
where $Z_2$ is the standard quark wave function renormalization
constant, while $m_{ph}$ is the mass to which a physical meaning
could be assigned. In other words, the quark always remains an
off-mass-shell object. Such an understanding (interpretation) of
quark confinement comes apparently from the Gribov's approach to
quark confinement \cite{10} and Preparata's massive quark model
(MQM) in which external quark legs were approximated by entire
functions \cite{11}. A quark propagator may or may not be an
entire function, but in any case the pole of the first order (like
the electron propagator has in QED) should disappear (see, for
example Refs. \cite{12,13,14} and references therein).

{\bf 2. The second sufficient condition}, formulated at the
hadronic (macroscopic) level, is the existence of the discrete
spectrum only (no continuum) in bound-states, in order to prevent
quarks to appear in asymptotic states. This condition comes
apparently from the 't Hooft's model for two-dimensional QCD with
large $N_c$ limit \cite{15} (see also Refs. \cite{12,16}).

This definition of quark confinement in the momentum space is
gauge invariant, flavor independent, i.e., valid for all types of
quarks (light or heavy), etc., and thus it is a general one. The
Wilson criterion of quark confinement formulated in the
configuration space - Area law \cite{17,18} is relevant only for
heavy quarks, as well as a linear rising potential between static
(heavy) quarks \cite{19}, also "seen" by lattice QCD \cite{20,21}.

At nonzero temperature and density, for example in the quark-gluon
plasma (QGP) \cite{22,23} (and references therein), the bound-states will be dissolved, so the second
sufficient condition does not work any more. However, the first
necessary condition remains always valid, of course. In other
words, by increasing temperature or density there is no way to put
quarks on the mass-shell. So that what is known as the
De-confinement phase transition in QGP is in fact the
De-hadronization phase transition. De-confinement is about the
liberation of the colored objects from the vacuum and not from the
bound-states. In the QCD ground-state there are many colored
objects such as quarks, gluons, instantons and may be something
else. Since color confinement is absolute and permanent, none of
these colored objects can appear in physical spectrum, and thus
De-confinement phase transition does not exist, in principle.

\section{Physical limits}

We introduced the renormalized mass gap in the relation (3.4),
defined its existence when the dimensionless UV regulating
parameter $\lambda$ goes to infinity, i.e., in the $\lambda
\rightarrow \infty$ limit. However, nothing was said about the
behavior of the coupling constant squared $g^2$ in this limit. In
principal, it may also depend on $\lambda$, becoming thus the
so-called "running" effective charge $g^2 \sim \alpha_s \equiv
\alpha_s(\lambda)$. In the general composition (3.4)

\begin{equation}
Z^{-1}(\lambda, \alpha_s(\lambda)) \Delta^2
(\lambda,\alpha_s(\lambda)),
\end{equation}
all the possible types of the effective charge behavior in the
$\lambda \rightarrow \infty$ limit should be considered
independently from each other (the dependence on other parameters
is not shown explicitly, as unimportant for further discussion).

1). If $\alpha_s(\lambda) \rightarrow \infty$ as $\lambda
\rightarrow \infty$, then one recovers the strong coupling regime.
Evidently, just this finite limit can be defined as the
renormalized mass gap (3.4), i.e., in fact

\begin{equation}
Z^{-1}(\lambda, \alpha_s(\lambda))\Delta^2 (\lambda,
\alpha_s(\lambda)) = \Delta^2_R, \quad \lambda \rightarrow \infty,
\quad \alpha_s(\lambda) \rightarrow \infty.
\end{equation}
Apparently, only this mass gap can be identified/related with/to
the Jaffe - Witten (JW) mass gap discussed in Ref. \cite{24} (see
section VII below as well).

2). If $\alpha_s(\lambda) \rightarrow c$ as $\lambda \rightarrow
\infty$, where $c$ is a finite constant, then it can be put unity,
not losing generality. This means that the effective charge
becomes unity, and this is only possible for the free gluon
propagator. But the free gluon propagator contains none of the
mass scale parameters, so in fact

\begin{equation}
Z^{-1}(\lambda, \alpha_s(\lambda))\Delta^2 (\lambda,
\alpha_s(\lambda)) =0, \quad \lambda \rightarrow \infty, \quad
\alpha_s(\lambda) \rightarrow 1.
\end{equation}

3). If $\alpha_s(\lambda) \rightarrow 0$ as $\lambda \rightarrow
\infty$, then one recovers the weak coupling regime. Evidently,
just this finite limit can be defined as $\Lambda^2_{QCD}$,
(however, see section VII below) i.e., in fact

\begin{equation}
Z^{-1}(\lambda, \alpha_s(\lambda))\Delta^2 (\lambda,
\alpha_s(\lambda)) = \Lambda^2_{QCD}, \quad \lambda \rightarrow
\infty, \quad \alpha_s(\lambda) \rightarrow 0.
\end{equation}
There is no doubt that the regularized mass gap may provide the
existence of the two different physical mass scale parameters
after the renormalization program is performed. Though these two
physical parameters show up explicitly at different regimes,
nevertheless, numerically they may not be very different, indeed,
as emphasized in Ref. \cite{1}. Our mass gap $\Delta^2_R$
determines the power-type deviation of the full gluon propagator
from the free one in the $q^2 \rightarrow 0$ limit. The region of
small $q^2$ is interesting for all the NP effects in QCD. This
once more emphasizes the close link between the behavior of QCD at
large distances and its INP dynamics. At the same time, the
asymptotic QCD scale parameter $\Lambda^2_{QCD}$ determines much
more weaker logarithmic deviation of the full gluon propagator
from the free one in the $q^2 \rightarrow \infty$ limit (see
subsection A below and Refs. \cite{3,4,5,6}). Then an interesting
question arises within our approach. How does exactly the
regularized mass gap provide the appearance of $\Lambda^2_{QCD}$
under the PT logarithms? The problem is that in the full gluon
propagator the regularized mass gap contribution is linearly
suppressed in comparison with the logarithmical divergent term in
the PT $q^2 \rightarrow \infty$ limit (see Refs. \cite{1,2}).

\subsection{Asymptotic freedom}

In the PT effective charge $d^{PT}(q^2) = [1 +\Pi^s(q^2; d^{PT})]^{-1}$ the invariant
function $\Pi^s(q^2; d^{PT})$ can be only logarithmical divergent \cite{1,2} in the PT
$q^2 \rightarrow \infty$ limit at any $d$, in particular at $d=d^{PT}$. So putting for
further convenience $d^{PT}(q^2) = \alpha_s(q^2; \Lambda^2) /
\alpha_s(\lambda)$ in this relation, one obtains

\begin{equation}
\alpha_s(q^2; \Lambda^2) = { \alpha_s(\lambda) \over 1 + b
\alpha_s(\lambda) \ln (q^2 / \Lambda^2)},
\end{equation}
and $b$ is the standard color group factor. This expression
represents the summation of the so-called main PT logarithms in
powers of $\alpha_s(\lambda)$. However, nothing should depend on
$\Lambda$ (and hence on $\lambda$) when they go to infinity in
order to recover the finite effective charge in this limit. To
show explicitly that this finite limit exists, let us formally
write

\begin{equation}
\Lambda^2 = f(\lambda)\Delta^2 (\lambda, \alpha_s(\lambda)),
\end{equation}
which is always valid, since $f(\lambda)$ is, in general, an
arbitrary dimensionless function. In this connection, let us again remind that in order to get the expression (4.5)
from the full effective charge $d(q^2) = [1 +\Pi^s(q^2; d) + c(d) (\Delta^2 / q^2)]^{-1}$
\cite{1,2} the mass gap contribution $\Delta^2 / q^2$ is only asymptotically suppressed in the $q^2 \rightarrow \infty$ limit. In other words, we distinguish between the asymptotic suppression of the mass gap contribution
$\Delta^2 / q^2$ and the formal PT $\Delta^2 =0$ limit. So the mass gap $ \Delta^2 = \Delta^2 (\lambda, \alpha_s(\lambda))$ itself here is not put identically zero and hence the relation (4.6) makes sense.
On account of the relation (4.4), it becomes

\begin{equation}
\Lambda^2 = f(\lambda) Z (\lambda, \alpha_s(\lambda))
\Lambda^2_{QCD}, \quad \lambda \rightarrow \infty, \quad
\alpha_s(\lambda) \rightarrow 0.
\end{equation}
Substituting it into the expression (4.5) and doing some algebra,
one obtains

\begin{equation}
\alpha_s(q^2) = { \alpha_s \over 1 + b \alpha_s \ln (q^2 /
\Lambda^2_{QCD})},
\end{equation}
if and only if

\begin{equation}
\alpha_s = { \alpha_s (\lambda) \over 1 - b \alpha_s (\lambda)
\ln(fZ)}, \quad \lambda \rightarrow \infty, \quad
\alpha_s(\lambda) \rightarrow 0.
\end{equation}
exists and is finite in the above shown limits. Here we introduce
the short hand notations $f \equiv f(\lambda)$ and $Z \equiv
Z(\lambda, \alpha_s(\lambda))$, for simplicity. Evidently, the
finite $\alpha_s$ can be identified with the fine-structure
constant of the strong interactions, calculated at some fixed scale.
It is worth emphasizing that the existence and finiteness of
$\alpha_s$ is due to the product $(fZ)$. Indeed, from Eq.~(4.9) it
follows

\begin{equation}
\ln(fZ) = {\alpha_s - \alpha_s (\lambda) \over \alpha_s b \alpha_s
(\lambda)} \rightarrow { 1 \over b \alpha_s (\lambda)}, \quad
\lambda \rightarrow \infty, \quad \alpha_s(\lambda) \rightarrow 0,
\end{equation}
which means that $(fZ)= \exp (1 /b \alpha_s (\lambda)$ in the
above shown limits. Substituting this into the relation (4.7) it
becomes

\begin{equation}
\lim_{(\Lambda, \lambda) \rightarrow \infty} \Lambda^2 \exp \Bigl(
- {1 \over b \alpha_s(\lambda)} \Bigr) = \Lambda^2_{QCD}, \quad
\alpha_s(\lambda) \rightarrow 0,
\end{equation}
which is the finite limit of the renormalization group equations
solution \cite{25}.

At very large $q^2$ from Eq.~(4.8) one recovers

\begin{equation}
\alpha_s(q^2) = { 1 \over  b \ln (q^2 / \Lambda^2_{QCD})},
\end{equation}
which is nothing but asymptotic freedom (AF) famous formula if
$b>0$ \cite{3,4,5,6,22}. In QCD with three colors and sixth
flavors this is so, indeed. For the pure Yang-Mills (YM) fields $b
= 11 / 4 \pi > 0$ always. Let us underline that in the expressions
(4.8) and (4.12) $q^2$ is always big enough, so it cannot go below
$\Lambda^2_{QCD}$. We have shown explicitly the AF behavior of QCD
at short distances ($q^2 \rightarrow \infty$), not using the
renormalization group equations and their solutions (i.e., we need
no expansion in powers of the coupling constant for the
corresponding $\beta$-function) \cite{3,4,5,6,25}. The regularized
mass gap is suppressed in the $q^2 \rightarrow \infty$ limit, as
it has been mentioned above. From Eq.~(4.4) and our
consideration in this subsection it follows, nevertheless, that
the regularized mass gap in the $\lambda \rightarrow \infty$ limit
provides the existence of the asymptotic QCD  scale parameter
$\Lambda^2_{QCD}$ as well.

There is no relation between the renormalized mass gap
$\Delta^2_R$ (4.2) and the asymptotic scale parameter
$\Lambda^2_{QCD}$ (4.4), since they show up explicitly at
different regimes. They are different scales, indeed, responsible
for different NP and non-trivial PT dynamics in QCD, though
numerically they may not be very different, as underlined above.
However, originally they have been generated in the region of
small $q^2$. In Ref. \cite{25} it has been noticed that being
numerically a few hundred $MeV$ only, $\Lambda^2_{QCD}$ cannot
survive in the $q^2 \rightarrow \infty$ limit. So none of the
finite mass scale parameters can be determined by the PT QCD. They
should come from the region of small $q^2$, being thus NP by
origin and surviving the renormalizatin program (i.e, the removal
of $\lambda$ in the $\lambda \rightarrow \infty$ limit), as was
just demonstrated above.

Concluding, all this can be a manifestation that "the problems
encountered in perturbation theory are not mere mathematical
artifacts but rather signify deep properties of the full theory"
\cite{26}. The message that we are trying to convey is that the
INP dynamical structure of the full gluon propagator indicates the
existence of its nontrivial PT one and the other way around.

\section{Confining potential}

After discussing some important aspects of the general MR program
for the INP gluon propagator using the DRM and the DT, it makes
sense to go back to the initial expansion for the INP gluon
propagator (2.1). A new surprising feature of this solution is
that its structure at zero ($q^2 \rightarrow 0$)
can be again determined by its $ \Delta^2 /(q^2)^2$ term only. To show this
explicitly, let us begin with the theorem from the theory of
functions of complex variable \cite{27}, which is an extremely
useful for the explanation of the behavior of our solution (2.1)
in the deep IR $q^2 \rightarrow 0$ limit.

The function $L(q^2)$ is defined by its Laurent expansion (2.2),
and thus it has an isolated essentially singular point at $q^2=0$.
Its behavior in the neighborhood of this point is regulated by the
Weierstrass-Sokhatsky-Casorati (WSC) theorem (see appendix B)
which tells that

\begin{equation}
\lim_{n \rightarrow \infty}L(q^2_n) = Y, \quad q^2_n
\rightarrow 0,
\end{equation}
where $Y$ is any complex number, and ${q^2_n}$ is a sequence
of points ${q^2_1, q^2_2 ... q^2_n ...}$ along which $q^2$ goes to
zero, and for which the above-displayed limit always exists. Of
course, $Y$ remains arbitrary (it depends on the chosen
sequence of points ${q^2_n}$), but, in general, it depends on the
same set of parameters as the residues, i.e., $Y \equiv Y(\lambda, \alpha, \xi, g^2)$.
This theorem thus allows one to replace the Laurent expansion $L(q^2)$ by $Y$ when $q^2
\rightarrow 0$ independently from all other test functions in the
corresponding skeleton loop integrands, i.e.,

\begin{equation}
L(q^2) \rightarrow Y(\lambda, \alpha, \xi, g^2), \quad q^2 \rightarrow 0.
\end{equation}

Our consideration in this section up to this point is necessarily
formal, since the mass gap remains unrenormalized. So far it has
been only regularized, i.e., $\Delta^2 \equiv \Delta^2(\lambda,
\alpha, \xi, g^2)$. The renormalization of the mass gap can be
proceed as follows. Due to the above-formulated WSC theorem, the
full gluon propagator (2.1) effectively becomes

\begin{equation}
D_{\mu\nu}(q) \equiv D_{\mu\nu}(q; \Delta^2) = i T_{\mu\nu}(q) { 1 \over (q^2)^2}
Y(\lambda, \alpha, \xi, g^2) \Delta^2 (\lambda, \alpha, \xi, g^2), \quad q^2 \rightarrow 0,
\end{equation}
so just the $\Delta^2 (q^2)^{-2}$ structure of the full gluon
propagator (2.1) is all that matters, indeed. Let us now define
the renormalized (R) mass gap as follows:

\begin{equation}
\Delta^2_R = Y(\lambda, \alpha, \xi, g^2) \Delta^2 (\lambda, \alpha, \xi, g^2),
\end{equation}
so that we consider $Y(\lambda, \alpha, \xi, g^2)$ as the MR constant for the mass gap, and $\Delta^2_R$ is the physical mass gap within our approach. Precisely this quantity should be positive, finite,
gauge-independent, etc., it should exist when $\lambda \rightarrow \infty$ and $\alpha \rightarrow 0$.Due to the WSC theorem, we can always choose such $Y = Y_1 \times Y_2 \times...$ in order to satisfy all the necessary requirements (each $Y_n$ will depend on its own sequence of points along which $q^2 \rightarrow 0$, the so-called subsequences).
Numerically it should be identified with the mass gap $\bar \Delta^2_R$ introduced by the relation (3.4) and described in section III before subsection A (see also the relation (5.8) below).

Thus the full gluon propagator relevant for the INP QCD dynamics
becomes

\begin{equation}
D_{\mu\nu}(q; \Delta^2_R) = i T_{\mu\nu}(q) { \Delta^2_R \over
(q^2)^2}.
\end{equation}
It is possible to say that because of the WSC theorem  there always exists such a sequence
of points along which $q^2 \rightarrow 0$ that the Laurent
expansion (2.1)-(2.2) effectively converges to the function (5.5) in the whole
$q^2$-momentum plane (containing both points $q^2=0$ and $q^2=
\infty$) after the renormalization of the mass gap is performed.
But the region of small $q^2$ is of special interest. In this
region the confinement dynamics begin to play a dominant role.
However, it is worth emphasizing that the solution (5.5) is not
the IR asymptotic ($q^2 \rightarrow 0$) of the initial Laurent
expansion (2.1)-(2.2), since $(q^2)^{-2}$ term is not a leading
one in this limit, it is a leading one just in the opposite $q^2 \rightarrow \infty$ limit.

The severe IR singularity
$(q^2)^{-2}$, which only one is present in the relevant gluon
propagator (5.5), is the first NP IR singularity possible in
four-dimensional QCD. A special regularization expansion is to be
used in order to deal with it, since the standard methods fail in
this case. As mentioned above, it should be correctly treated
within the DT \cite{7} complemented by the DRM \cite{8}. If $q^2$
is an independent skeleton loop variable, then the corresponding
dimensional regularization of this singularity is given by the
expansion (3.2) at $k=0$, namely

\begin{equation}
(q^2)^{- 2} = { 1 \over \epsilon} \Bigr[ a(0)\delta^4(q) +
O_0(\epsilon) \Bigl], \quad \epsilon \rightarrow 0^+,
\end{equation}
and $a(0) = \pi^2$. Due to the $\delta^4(q)$ function in the
residue of this expansion, all the test functions which appear
under corresponding skeleton loop integrals should be finally
replaced by their expression at $q=0$. So the dimensionally
regularized expansion for the gluon propagator (5.5) becomes

\begin{equation}
D_{\mu\nu}(q; \Delta^2_R) = i T_{\mu\nu}(q) \Delta^2_R \times {1
\over \epsilon} \Bigr[a(0) \delta^4(q) + O(\epsilon) \Bigl], \quad
\epsilon \rightarrow 0^+,
\end{equation}
where we put $O_0(\epsilon) = O(\epsilon)$, for simplicity. As emphasized above, in the
presence of the IR regularization parameter $\epsilon$ all the
Green's functions and parameters become, in general, dependent on it. The
only way to remove the pole $ 1 / \epsilon$ from the relevant
gluon propagator (5.7) is to define the IR renormalized mass gap
as follows:

\begin{equation}
\Delta^2_R = X(\epsilon) \bar{\Delta}^2_R = \epsilon \times
\bar{\Delta}^2_R, \quad \epsilon \rightarrow 0^+,
\end{equation}
where $X(\epsilon) = \epsilon$ is the IR MR constant for the mass
gap. Let us remind that contrary to $\Delta^2_R$ its IR renormalized counterpart
$\bar{\Delta}^2_R$ exist as $\epsilon \rightarrow 0^+$.
In both expressions for the mass gap the dependence on $\epsilon$ is assumed but
not shown explicitly. So we distinguish between both mass gaps only by the dependence on $\epsilon$.

On account of the relation (5.8), the dimensionally regularized
expansion (5.7) finally becomes

\begin{equation}
D_{\mu\nu}(q; \bar{\Delta}^2_R) = i T_{\mu\nu}(q) \bar{\Delta}^2_R
a(0)\delta^4(q) + O(\epsilon), \quad \epsilon \rightarrow 0^+.
\end{equation}
Evidently, after performing the renormalization program (i.e.,
going to the IR renormalized quantities), the terms of the order
$O(\epsilon)$ can be omitted from the consideration.

The renormalization of the mass gap automatically IR renormalizes
the relevant gluon propagator (5.5), so that

\begin{equation}
D_{\mu\nu}(q; \bar \Delta^2_R) = \epsilon \times i T_{\mu\nu}(q)
{\bar{\Delta}^2_R \over (q^2)^2}, \quad \epsilon \rightarrow 0^+,
\end{equation}
in complete agreement with the expansion (3.9) at $k=0$ with
putting there $\bar \Omega^R_0=1$. In order to achieve the agreement with the general
renormalization program performed in section III it is necessary to put
$Y^{-1}(\lambda, \epsilon) X(\epsilon) = Z (\lambda, \epsilon)$ and hence
$\Phi_0(\lambda; \epsilon) = \bar{\Omega}_0^R Y(\lambda, \epsilon) = Y(\lambda, \epsilon)$, on account of the
convergence condition (3.6) at $k=0$ and the relation (5.8).

There is no doubt left that the WSC theorem somehow underlines the
importance of the simplest NP IR singularity $ 1 / (q^2)^2$
possible in four-dimensional QCD, while all other may be suppressed
in the deep IR ($q^2 \rightarrow 0$) region due to this theorem.
However, from the general consideration in sections III and IV the
suppression mechanism is not seen, at least at this stage. In principle, one can develop
the formal PT series in powers of the mass gap. Then the first
nontrivial approximation is just the above-mentioned singularity $
\Delta^2_R / (q^2)^2$, which is only one to be started with,
anyway. It is well known that the expression (5.5) leads to the
linear rising potential between heavy quarks \cite{19} also "seen"
by lattice QCD \cite{20,21}. That is why we call it the confining
potential. Just it will be used for the derivation of the system
of equations determining the confining quark propagator (for
preliminary consideration see Ref. \cite{28} and references
therein). Evidently, what we have pointed out in subsection A of
section III for the general case is valid for the confining
potential (5.5) as well.

\subsection{The renormalized "running" effective charge}

It is instructive to find explicitly the corresponding
$\beta$-function. From Eq.~(5.5) it follows that the corresponding
Lorentz structure, which is nothing but the corresponding
effective charge ("running"), in terms of the renormalized mass
gap is

\begin{equation}
d(q^2; \Delta^2_R) \equiv \alpha_s(q^2; \Delta^2_R) = {\Delta^2_R
\over q^2},
\end{equation}
and this does not depend whether the gluon momentum is independent
loop variable or not. Then from the renormalization group equation
for the renormalized effective charge, which determines the $\beta$-function,

\begin{equation}
q^2 {d \alpha_s(q^2; \Delta^2_R) \over dq^2} = \beta(\alpha_s(q^2;
\Delta^2_R)),
\end{equation}
it simply follows that

\begin{equation}
\beta(\alpha_s(q^2; \Delta^2_R))= - \alpha_s(q^2; \Delta^2_R) = -
{\Delta^2_R \over q^2}.
\end{equation}
Thus, one can conclude that the corresponding $\beta$-function as
a function of its argument is always in the domain of attraction
(i.e., negative). So it has no IR stable fixed point indeed as it
is required for the confining theory \cite{3}. Just these
expressions for the $\beta$-function and the running effective
charge should be used for the calculation of the truly NP
quantities in the YM theory, such as the gluon condensate, the
gluon part of the Bag constant \cite{29,30,31}, etc. In phenomenology
for these purposes we need to know the ratio $\beta(q^2) /
\alpha_s(q^2)$ in the $q^2 \rightarrow 0$ limit, so it is always

\begin{equation}
{\beta(q^2) \over \alpha_s(q^2)} \equiv {\beta(\alpha_s(q^2;
\Delta^2_R)) \over \alpha_s(q^2; \Delta^2_R)} = - 1
\end{equation}
within our approach to low-energy QCD, that's INP QCD (don't
forget that the replacement $\alpha_s(q^2) \rightarrow
\alpha_s^{INP}(q^2)$ and $\beta(q^2) \rightarrow \beta^{INP}(q^2)$
is assumed \cite{2}).

Concluding, let us note if the renormalized effective charge (5.11) is to be
multiplied by the additional powers of $(q^2)^{-2-k}, \ k=0,1,2,3,...$, then this product should be treated
as the full gluon propagator itself, described above. The only difference between them becomes the unimportant
tensor structure $T_{\mu\nu}(q)$.

\section{ Discussion}

Thus the QCD vacuum is really beset with severe IR
singularities, which have been summarized (accumulated) into the
INP gluon propagator (2.1). There is no doubt that the purely
transversal severely singular virtual gluon field configurations
play important role in the dynamical and topological structure of
the QCD ground state, leading thus to the general zero
momentum modes enhancement (ZMME) effect there reflected in the
INP gluon propagator. Evidently, the  ZMME (or simply ZME)
mechanism of confinement (see our previous papers \cite{32,33}
and references therein as well) is nothing but the well forgotten IR
slavery (IRS) one, which can be equivalently referred to as a
strong coupling regime \cite{3,16}.

Indeed, at the very beginning of QCD it was expressed a general
idea \cite{16,34,35,36,37,38,39,40} that the quantum excitations
of the IR degrees of freedom, because of self-interaction of
massless gluons in the QCD vacuum, made it only possible to
understand confinement, dynamical (spontaneous) breakdown of
chiral symmetry and other NP effects. In other words, the
importance of the deep IR structure of the QCD vacuum has
been emphasized as well as its relevance to the above-mentioned NP
effects and the other way around. This development was stopped by
the wide-spread wrong opinion that severe IR singularities cannot
be put under control. Here we have explicitly shown (see also our
recent papers \cite{9,41} and references therein) that the
adequate mathematical theory for quantum YM gauge theory is the DT
(the theory of generalized functions) \cite{7}, complemented by
the DRM \cite{8}. Together with the theory of functions of complex
variable \cite{27} they provide a correct treatment of these severe
IR singularities without any problems. Thus, we come back to the
old idea but on a new basis that is why it becomes new ("new is
well forgotten old"). In other words, we put the IRS mechanism of
color confinement on a firm mathematical ground. This makes it
possible to analytically formulate the gluon confinement criterion
in a gauge invariant way for the first time.

The confining potential (5.5) in the different approximations and
gauges has been earlier obtained and investigated in many papers
(see, for example Ref. \cite{42} and references therein). We have
confirmed and thus revitalized these investigations, in which this
behavior has been obtained as an IR asymptotic solution to the
gluon SD equation. However, let us emphasize once more that due to
the WSC theorem the confining potential (5.5) is not the IR
asymptotic of the initial Laurent expansion (2.1)-(2.2). Moreover,
the whole INP gluon propagator (2.1) effectively converges to the
confining potential (5.5) after the renormalization of the mass
gap is performed. The WSC theorem clearly shows that the
$\Delta^2_R /(q^2)^2$ structure is only important, while all other
terms in the INP gluon propagator (2.1) are suppressed (though
each next term in the expansion (2.2) is more singular in the IR
than the previous one).

In the presence of the mass gap the coupling constant plays no
role. This is also an evidence of the "dimensional transmutation",
$g^2 \rightarrow \Delta^2(\lambda, \alpha, \xi, g^2)$
\cite{3,43,44}, which occurs whenever a massless theory acquires
masses dynamically. It is a general feature of spontaneous
symmetry breaking in field theories. In our case, the color gauge
symmetry is broken at the level of the full gluon self-energy,
while maintaining at the level of the full gluon propagator, which
is more important \cite{1} (something like a "self-consistent
violation" or a "hidden violation" of the color gauge symmetry).
In the massive solution \cite{2} the mass gap transforms further into the
effective gluon mass, i.e., $g^2 \rightarrow \Delta^2(\lambda,
\alpha, \xi, g^2) \rightarrow m_g(\xi)$. Nevertheless, it remains
gauge-dependent (i.e., not physical) even after the corresponding
full gluon propagator is renormalized. Within the INP QCD the mass
gap transforms further into the physical mass gap, i.e., $g^2
\rightarrow \Delta^2(\lambda, \alpha, \xi, g^2) \rightarrow
\Delta^2_R$, and the gluons remain massless in a gauge invariant
way (this paper). Let us note that the renormalization of the mass gap automatically renormalizes
the INP QCD gluon propagator as well.

\section{Conclusions}

Let us denote the version of our mass gap $\bar \Delta^2_R$ which will
appear in the $S$-matrix elements for the corresponding physical
quantities/processes in low-energy QCD as $\Lambda^2_{INP}$ (in
principle, they may be slightly different from each other,
indeed). Then a symbolic relation between it and the initial mass
gap $\Delta^2(\lambda, \alpha_s(\lambda))$ and $\Lambda^2_{PT}$
instead of $\Lambda^2_{QCD}$ (for reason see discussion below)
could be written as follows:

\begin{equation}
\Lambda^2_{INP} \longleftarrow^{\infty \leftarrow
\alpha_s(\lambda)}_{\infty \leftarrow \lambda} \ \Delta^2(\lambda,
\alpha_s(\lambda)) \ { }^{\alpha_s(\lambda) \rightarrow
0}_{\lambda \rightarrow \infty} \longrightarrow  \ \Lambda^2_{PT},
\end{equation}
which summarizes our main results in this investigation. QCD as a
quantum gauge field theory, describing the interactions of never
seen colored objects (the gluons and quarks), cannot have the
physical mass gap. In other words, this is a theory which describes the behavior of
the colored objects in the vacuum. In QCD mass gap may only appear in the
way described in our previous publication \cite{1}, that's
$\Delta^2(\lambda, \alpha_s(\lambda))$ in Eq.~(7.1). In order to
become the theory of the strong interactions it should undergo the
two phase transitions; in the weak and strong coupling regimes. In
the first case it becomes the PT QCD which describes all the
high-energy phenomena in the strong interactions from first
principles (AF, scale violation, hard processes, etc.). It has its
own physical mass gap which we denote as $\Lambda^2_{PT}$ in
Eq.~(7.1). In the second case it becomes the INP QCD which
describes all the low-energy phenomena in the strong interactions
from first principles (those includes first of all color
confinement, dynamical breakdown of chiral symmetry, bound-states,
etc.). It has its own physical mass gap, that's $\Lambda^2_{INP}$
in Eq.~(7.1).

In this connection, a few things should be made perfectly clear.
First of all, let us underline that the PT QCD and the INP QCD are
not effective theories, as pointed out above both theories are
fundamental ones. Secondly, such a quantity as $\Lambda^2_{QCD}$
does not exist at all, since QCD itself cannot have a physical
limit. In order to avoid any confusion the corresponding scale is
better to denote as $\Lambda^2_{PT}$, since just the PT QCD is
responsible for all the high-energy phenomena in the strong
interactions. Thus similarly the relation (7.1), the following
symbolic relation makes sense

\begin{equation}
INP \ QCD \ \Longleftarrow \ QCD \ \Longrightarrow \ PT \ QCD,
\end{equation}
so that at the fundamental (quark-gluon) level the PT QCD is AF,
while the INP QCD confines gluons. In the subsequent papers we
will show that this theory will confine quarks, as well as will
explain the spontaneous breakdown of chiral symmetry. Both
theories have their own mass gaps $\Lambda^2_{INP}$ and
$\Lambda^2_{PT}$, which are solely responsible for the large-
and short-scale structures of the QCD ground state,
respectively.

A few years ago Jaffe and Witten have formulated the following
theorem \cite{21}:

\vspace{3mm}

{\bf Yang-Mills Existence And Mass Gap:} Prove that for any
compact simple gauge group $G$, quantum Yang-Mills theory on
$\bf{R}^4$ exists and has a mass gap $\Delta > 0$.

\vspace{3mm}

Of course, to prove the existence of the YM theory with compact
simple gauge group $G$ is a formidable task yet. It is rather a
mathematical than a physical problem. However, from the JW
presentation of their theorem it clearly follows that their mass
gap should be identified with our mass gap $\Lambda^2_{INP}$. At
the same time, we have argued above that QCD itself cannot have a physical mass gap.
It has a mass gap which is only regularized, i.e., $\Delta^2
\equiv \Delta^2(\lambda, \alpha_s(\lambda))$, and therefore there
is no guarantee that it is positive. It cannot be related directly
to any of physical quantities/processes. Let us also remind that QCD cannot confine free gluons \cite{1,2}.
As actual theory of the strong interactions the two different faces of QCD come into the play:
the PT QCD for high-energy physics and INP QCD for low-energy physics, which confines "dressed" gluons, while free gluons do not exist in this theory. The corresponding mass gaps have now physical meanings (they are
finite, positive, gauge-invariant, etc.).

Our basic result obtained in the previous works
\cite{1,2} and in this paper can be jointly formulated as follows:

\vspace{3mm}

{\bf Mass Gap Existence And Gluon Confinement:} If quantum Yang-Mills theory with
compact simple gauge group $G=SU(3)$ exists on $\bf{R}^4$,
then undergoing the phase transition in the strong coupling regime it becomes INP QCD,
which has a physical mass gap and confines gluons.

\vspace{3mm}

Some important features of the INP QCD are:

1. Its full gluon propagator (2.1) converges to the expression (5.5) after the
renormalization of the mass gap is performed. This expression is effectively valid 
in the whole $q^2$-momentum plane.

2. It has a physical mass gap.

3. It confines "dressed" gluons in asymptotic states.

4. It has no free gluons at all due to the subtraction method proposed and formulated in
   our previous works \cite{1,2}.

Some other interesting features of this theory may be established after the explicit including
of the quark degrees of freedom into its formalism in the next our papers. However, first in the subsequent paper
we will show how the QCD vacuum structure is to be investigated within the YM version of the INP QCD.

\vspace{3mm}

Concluding, a few remarks are in order. In our previous work
\cite{1}, we have explicitly shown that the full photon propagator
in quantum electrodynamics (QED) has only the PT-type IR
singularity, $1/q^2$. This is in agreement with the cluster
property of the Wightman functions \cite{45}, that's correlation
functions of observables. In QCD the explicit presence of the
regularized mass gap, which are necessarily accompanied by severe
IR singularities $(q^2)^{-2-k}, \ k=0,1,2,3...$, apparently, will
violate this property. In turn, this validates the Strocchi
theorem \cite{46}, which allows such a severely singular behavior
of the full gluon propagator in QCD. However, this is not a
problem, since QCD has no physical observables. In PT
QCD, which gluon propagator is as much singular as $1/q^2$ only,
the cluster property will not be violated. On the other hand, in
INP QCD with a such singular behavior of the relevant gluon
propagator (5.5) the situation with the Wightman functions is not
clear. It can be clarified only after the solution of the color
confinement problem, and a realistic calculations of the various
physical observables within INP QCD. At the fundamental
quark-gluon level only these remarks make sense about the
correlation between the structure of the corresponding gluon propagator
and the properties of the Wightman functions.

\begin{acknowledgments}

Support by HAS-JINR grant (P. Levai) is to be acknowledged. The
author is grateful to P. Forg\'{a}cs, J. Nyiri, T. Bir\'{o}, C.
Wilkin, A. Luk\'{a}cz, M. Vas\'{u}ht, G. Barnaf\"{o}ldi and especially
to A.V. Kouzushin for useful discussions, remarks and help.

\end{acknowledgments}

\appendix

\section{Infrared dimensional regularization within the distribution theory}

\subsection{The DRM in the PT}

As repeatedly  emphasized in our previous works \cite{1,2}, the
mass gap in nothing but the re-defined skeleton tadpole term, or
it can be reduced to the skeleton tadpole term itself. Also, it
has been explicitly shown that there is no such regularization
scheme (preserving or not gauge invariance) in which the
transversality condition for the full gluon self-energy could be
satisfied unless the constant skeleton tadpole term

\begin{equation}
\Pi^t_{\rho\sigma}(D) \equiv \Pi_t(D) \equiv \Delta_t^2(D) =  g^2
\int {i d^4 q_1 \over (2 \pi)^4} T^0_4 D(q_1),
\end{equation}
is to be disregarded from the very beginning, i.e., put formally
zero everywhere. Here $T^0_4$ is the four-gluon point-like vertex,
and $g^2$ is the dimensionless coupling constant squared. We omit
the tensor and color indices in this integral, as unimportant for
further discussion. It is nothing else but the quadratically
divergent in the PT constant, so it is assumed to be regularized.
The mass gap is not survived in the PT $q^2 \rightarrow \infty$
limit \cite{1}, however, in the PT there are still problems with
such kind of integrals.

In the PT, when the first non-trivial approximation for the full
gluon propagator $D = D(q)$ is the free one $D_0=D_0(q)$, the
constant tadpole term is to be simply discarded, i.e., to be put
formally zero within the DRM \cite{8}, so that
$\Pi^t_{\rho\sigma}(D_0) =\delta_{\rho\sigma} \Delta^2_t(D_0) =
0$. However, even in the DRM this is not an exact result, but
rather an embarrassing prescription, as pointed out in Ref.
\cite{5}. To show explicitly that there are still problems, as
mentioned above, it is instructive to substitute the first
iteration of the gluon SD equation into the previous expression
(A1). Symbolically it looks like $D(q) =  D_0(q) + D_0(q)i\Pi(q;
D)D(q) = D_0(q) + D_0(q)i\Pi(q; D_0)D_0(q) + ... = D_0(q) +
D^{(1)}(q) + ...$, where we omit all the indices and put $D_0
\equiv D^{(0)}$. Doing so, one obtains

\begin{eqnarray}
\Pi_t(D=D_0 + D^{(1)}+...) &=& \Pi_t(D_0) + \Pi_t(D^{(1)}) + ... =
\Pi_t(D_0) + g^2 \int {i d^4 q_1 \over (2 \pi)^4} T^0_4
[D_0(q_1)]^2 i \Pi(q_1; D_0) +...
\nonumber\\
&=& \Pi_t(D_0) + \Pi_t(D_0) g^2 \int {i^2 d^4 q_1 \over (2 \pi)^4}
T^0_4 [D_0(q_1)]^2 + g^2 \int {i^2 d^4 q_1 \over (2 \pi)^4} T^0_4
[D_0(q_1)]^2 \Pi^s(q_1; D_0) + ... \ . \nonumber\\
\end{eqnarray}
Here we introduce the subtraction as follows: $\Pi^s(q_1; D_0) =
\Pi(q_1; D_0) - \Pi(0; D_0)$, and  put $\Pi(0; D_0)= \Pi_t(D_0)$,
for simplicity, when the mass gap is to be reduced to the tadpole
term itself \cite{1}. In the second line of Eq. (A2) the first
integral is not only UV divergent but IR singular as well. If we
now omit the first term in accordance with the above-mentioned
prescription, the product of this integral and the tadpole term
$\Pi_t(D_0)$ remains, nevertheless, undetermined. Moreover, the
structure of the second integral in this line is much more
complicated than in the divergent constant integral $\Pi_t(D_0)$
in Eq.~(A1). All this reflects the general problem that such kind
of massless integrals

\begin{equation}
\int { d^d q \over (2 \pi)^d} {q_{\mu_1}...q_{\mu_p} \over
(q^2)^n}
\end{equation}
are ill defined, since there is no dimension where they are
meaningful. They are either IR singular or UV divergent, depending
on the relation between the numbers $d$, $p$ and $n$ \cite{5}.
This prescription clearly shows that the DRM, preserving gauge
invariance, nevertheless, is by itself not sufficient to provide
us insights into the correct treatment of the power-type IR
singularities shown in Eq.~(A3) (we will address this problem
below in subsection 2). Thus, one concludes that the tadpole term
(A1) $\Delta^2_t(D) \equiv \Delta^2_t(\lambda, \alpha; D)$ is, in
general, not zero.

However, in the PT we can adhere to the prescription that such
massless tadpole integrals can be discarded in the DRM \cite{5,8}.
As mentioned above, we have already shown \cite{1} that the mass
gap, in general, and the skeleton tadpole term, in particular, can
be neglected in the PT, indeed (not depending on whether $\lambda,
\alpha$ are to be introduced within the regularization scheme
preserving gauge invariance or not). In what follows we will show
how precisely the DRM \cite{8} should be correctly implemented
into the DT \cite{7} in order to control the power-type severe IR
singularities, which may appear not only in the PT series, but
mainly in the NP QCD as well (see Ref. \cite{2}).

\subsection{The DRM in the DT}

In general, all the Green's functions in QCD are generalized
functions, i.e., they are distributions. This is especially true
for the NP IR singularities of the full gluon propagator due to
the self-interaction of massless gluons in the QCD vacuum. They
present a rather broad and important class of functions with
algebraic singularities, i.e., functions with nonsummable
singularities at isolated points \cite{7} (at zero in our case).
Roughly speaking, this means that all relations involving
distributions should be considered under corresponding integrals,
taking into account the smoothness properties of the corresponding
space of test functions. Let us note in advance that the
space in which our generalized functions are continuous linear
functionals is $K$, that's the space of infinitely differentiable functions having compact support,
i.e., they are zero outside some finite region (different for each differentiable function) \cite{7}.

Let us consider the positively definite ($P>0$) squared
(quadratic) Euclidean form

\begin{equation}
P(q) = q_0^2 +  q_1^2 + q_2^2 + ... + q_{n-1}^2 = q^2,
\end{equation}
where $n$ is the number of the components. The generalized
function (distribution) $P^{\lambda}(q)$, where $\lambda$ being, in
general, an arbitrary complex number, is defined as

\begin{equation}
(P^{\lambda}, \varphi) = \int_{P>0}P^{\lambda}(q) \varphi(q) d^dq,
\end{equation}
where $\varphi(q)$ is the above-mentioned some test function.
At $Re \lambda \geq 0$ this integral is convergent and is an
analytic function of $\lambda$. Analytical continuation to the
region $Re \lambda < 0$ shows that it has a simple pole at points
\cite{7}

\begin{equation}
\lambda = - {n \over 2} - k, \quad k=0, 1, 2 ,3...
\end{equation}

In order to actually define the skeleton loop integrals in the deep IR domain, which
appear in the system of the SD equations, it is necessary to introduce the IR regularization
parameter $\epsilon$, defined as $d = n + 2 \epsilon, \ \epsilon
\rightarrow 0^+$ within the DRM \cite{8}, where $d$ is the dimension of the loop integral
(see Eq.~(A5)). As a result, all the Green's functions and "bare" parameters should be
regularized with respect to $\epsilon$ which should be set to zero
at the end of the computations. The structure of the NP IR
singularities is then determined (when $n$ is even number) as follows \cite{7}:

\begin{equation}
(q^2)^{\lambda} = { C_{-1}^{(k)} \over \lambda +(d/2) + k} +
finite \ terms,
\end{equation}
where the residue is

\begin{equation}
 C_{-1}^{(k)} = { \pi^{n/2} \over 2^{2k} k! \Gamma ((n/2) + k) } \times
L^k \delta^n (q)
\end{equation}
with $L = (\partial^2 / \partial q^2_0) + (\partial^2 /
\partial q^2_1) + ... + (\partial^2 / \partial q^2_{n-1})$.

Thus the regularization of the NP IR singularities (A5), on
account of (A6), is nothing but the whole expansion in the
corresponding powers of $\epsilon$ and not the separate term(s).
Let us underline its most remarkable feature. The order of
singularity does not depend on $\lambda$, $n$ and $k$. In terms of
the IR regularization parameter $\epsilon$ it is always a simple
pole $1/ \epsilon$. This means that all power terms in Eq. (A7)
will have the same singularity, i.e.,

\begin{equation}
(q^2)^{- {n \over 2} - k } = { 1 \over \epsilon} C_{-1}^{(k)} +
finite \ terms, \quad \epsilon \rightarrow 0^+,
\end{equation}
where we can put $d=n$ now (i.e., after introducing this
expansion). By "$finite \ terms$" here and everywhere a number of
necessary subtractions under corresponding integrals is understood
\cite{7}. However, the residue at a pole will be drastically
changed from one power singularity to another. This means
different solutions to the whole system of the SD equations for
different set of numbers $\lambda$ and $k$. Different solutions
mean, in their turn, different vacua. In this picture different
vacua are to be labeled by the two independent numbers: the
exponent $\lambda$ and $k$. At a given number of $d(=n)$ the
exponent $\lambda$ is always negative being integer if $d(=n)$ is
an even number or fractional if $d(=n)$ is an odd number. The
number $k$ is always integer and positive and precisely it
determines the corresponding residue at a simple pole, see Eq.
(A9). It would not be surprising if these numbers were somehow
related to the nontrivial topology of the QCD vacuum in any
dimensions.

It is worth emphasizing that the structure of severe IR
singularities in Euclidean space is much simpler than in Minkowski
space, where kinematical (unphysical) singularities due to the
light cone also exist \cite{3,7,47} (in this connection let us
remind that in Euclidean metrics $q^2 = 0$ implies $q_i=0$ and
vice-versa, while in Minkowski metrics this is not so). In this
case it is rather difficult to untangle them correctly from the
dynamical singularities, the only ones which are important for the
calculation of any physical observable. Also, the consideration is
much more complicated in the configuration space \cite{7}. That is
why we always prefer to work in the momentum space (where
propagators do not depend explicitly on the number of dimensions)
with Euclidean signature. We also prefer to work in the covariant
gauges in order to avoid peculiarities of the non-covariant gauges
\cite{3,48,49}, for example, how to untangle the gauge pole from
the dynamical one.

In principle, none of the regularization schemes (how to introduce
the IR regularization parameter in order to parameterize the NP IR
divergences and thus to put them under control) should be
introduced by hand. First of all, it should be well defined.
Secondly, it should be compatible with the DT \cite{7}. The DRM
\cite{8} is well defined, and here we have shown how it should be
introduced into the DT (complemented by the number of
subtractions, if necessary). Though the so-called $\pm i\epsilon$
regularization is formally equivalent to the regularization used
in our paper (see again Ref. \cite{7}), nevertheless, it is rather
inconvenient for practical use. Especially this is true for the
gauge-field propagators, which are substantially modified due to
the response of the vacuum (the $\pm i\epsilon$ prescription is
designated for and is applicable only to the theories with the PT
vacua, indeed \cite{10,50}). Other regularization schemes are also
available, for example, such as analytical regularization used in
Ref. \cite{16} or the so-called Speer's regularization \cite{51}.
However, they should be compatible with the DT as emphasized
above. Anyway, not the regularization is important but the DT
itself. Just this theory provides an adequate mathematical
framework for the correct treatment of all the Green's functions
in QCD (apparently, for the first time the distribution nature of
the Green's functions in quantum field theory has been recognized
and used in Ref. \cite{52}).

The regularization of the NP IR singularities in QCD is determined
by the Laurent expansion (A9) at $n=4$ as follows:

\begin{equation}
(q^2)^{- 2 - k } = { 1 \over \epsilon} a(k)[\delta^4(q)]^{(k)} +
f.t. = { 1 \over \epsilon} \Bigr[ a(k)[\delta^4(q)]^{(k)} +
O_k(\epsilon) \Bigl], \quad \epsilon \rightarrow 0^+,
\end{equation}
where $a(k)= \pi^2 / 2^{2k} k! \Gamma(2+k)$ is a finite constant
depending only on $k$ and $[\delta^4(q)]^{(k)}$ represents the
$k$'s derivative of the $\delta$-function (see Eqs.~(A7) and
(A8)). We point out that after introducing this expansion
everywhere one can fix the number of dimensions, i.e., put $d=n=4$
for QCD without any further problems. Indeed there will be no
other severe IR singularities with respect to $\epsilon$ as it
goes to zero, but those explicitly shown in this expansion. Let us
underline that, while the initial expansion (2.1) is the Laurent
expansion in the inverse powers of the gluon momentum squared, the
regularization expansion (A10) is the Laurent expansion in powers
of $\epsilon$. This means that its regular part is as follows:
$f.t. =(q^2)^{- 2 - k }_{-} + \epsilon (q^2)^{- 2 - k }_{-} \ln
q^2 + O(\epsilon^2)$, where for the unimportant here definition of
the functional $(q^2)^{- 2 - k }_{-}$ see Ref. \cite{7}. These
terms, however, play no any role in the IRMR program which has
been discussed in section III. The dimensionally regularized
expansion (A10) takes place only in four-dimensional QCD with
Euclidean signature. In other dimensions and/or Minkowski
signature it is much more complicated as pointed out above. As it
follows from this expansion any power-type NP IR singularity,
including the simplest one at $k=0$, scales as $1 /\epsilon$ as it
goes to zero. Just this plays a crucial role in the IR
renormalization of the theory within our approach. Evidently, such
kind of the dimensionally regularized expansion (A10) does not
exist for the PT IR singularity, which is as much singular as
$(q^2)^{-1}$ only.

 In summary, first we have emphasized the
distribution nature of the NP IR singularities. Secondly, we have
explicitly shown how the DRM should be correctly and in a gauge
invariant way implemented into the DT. This makes it possible to
put severe IR singularities under firm mathematical control.

\section{The WSC theorem}

One of the main theorems in the theory of functions of complex
variable \cite{27} is the the above-mentioned WSC theorem. It
describes the behavior of meromorphic functions near essential
singularities.

\vspace{3mm}

{\bf Theorem (Weierstrass-Sokhatsky-Casorati).} If $z_0$ is an
essential singularity of the function $f(z)$, then for any complex
number $Z$ there exists the sequence of points $z_k \rightarrow
z_0$, such that

\begin{equation}
\lim_{k \rightarrow \infty} f(z_k) = Z.
\end{equation}

\vspace{3mm}

So this theorem tells us that the behavior of the
function $f(z)$ near its essential singularity $z_0$ is not
determined, i.e, in fact, it remains arbitrary. It depends on the
chosen sequence of points $z_k$ along which $z$ goes to zero,
that's $Z \equiv Z({z_k})$ (do not mixed this complex number with the
MR constant of the mass gap in the relation (3.4)).

Let us consider one classical example \cite{27}. The function

\begin{equation}
f(z) = e^{1 / z} = \sum_{n=0}^{\infty} { 1 \over n! z^n}
\end{equation}
has the above shown Laurent series about the essential singularity
at $z=0$, i.e., this Laurent expansion converges to the function
$f(z)$ everywhere apart from the point $z=0$. At the same time,
due to the WSC theorem the behavior of the function $f(z)$ near
$z=0$, and hence of the Laurent expansion itself, depends on the
chosen sequence of points ${z_k}$ along which $z \rightarrow 0$.
So let us proceed as follows:

{\bf (i).} If one chooses $z_k = 1 / k, \ k=1,2,3,...$, then

\begin{equation}
\lim_{k \rightarrow \infty} f(z_k) = \lim_{k \rightarrow \infty}
e^k = \infty.
\end{equation}

{\bf (ii).} If one chooses $z_k = - 1 / k, \ k=1,2,3,...$, then

\begin{equation}
\lim_{k \rightarrow \infty} f(z_k) = \lim_{k \rightarrow \infty}
e^{-k} = 0.
\end{equation}

{\bf (iii).} If one chooses $z_k = 1 / \ln A + 2k \pi i, \
k=0,1,2,3,...$, then

\begin{equation}
\lim_{k \rightarrow \infty} f(z_k) = \lim_{k \rightarrow \infty}
e^{\ln A + 2k \pi i} = A,
\end{equation}
where $A$ is some finite constant. Many other examples can be
found in Ref. \cite{27} and in other text-books on the theory of
functions of complex variable.

Concluding, let us make one important point perfectly clear. For
example, the function

\begin{equation}
f(z) = {z \over \sqrt{1 + z^2}} = \sum_{n=0}^{\infty} (-1)^n
{(2n)! \over 2^{2n} (n!)^2} { 1 \over z^{2n}}
\end{equation}
has no singularity at zero at all, while its Laurent expansion
always has an essential singularity at $z=0$, by definition. The
equality (B6) means that this Laurent expansion converges to the
above shown function $f(z)$ in the ring which excludes zero point.
Its region of convergence is $1 < |z| < \infty$, while the
behavior of any Laurent expansion near its essential singularity
is always governed by the WSC theorem, in particular of the
Laurent expansion shown in the right-hand-side of Eq.~(B6).

Another characteristic example is the function

\begin{equation}
f(z) = {z \over z - 1} = \sum_{n=0}^{\infty} \Bigl( { 1 \over z}
\Bigr)^n,
\end{equation}
which is nothing but geometric series. The Laurent expansion,
shown in the right-hand-side of this equation, converges to the
function $f(z)$ also in the region $1 < |z| < \infty$, while the
behavior of the Laurent expansion itself at $z \rightarrow 0$ is
again governed by the WSC theorem. The same is true for the
Laurent expansion in Eq.~(B2). It converges to the function
$\exp(1 / z)$ in the whole complex plane apart from the point $z=0$,
while near this point its behavior is uncertain, as it was
described above. The message we are trying to convey is that all
the equalities containing the Laurent expansions should be treated
carefully in accordance with the above-mentioned theorem (i.e.,
the region of convergence should be fixed clearly, otherwise the
equality can be incorrectly understood).

\end{document}